\documentclass[sigconf]{acmart}

\usepackage{listings} 
\usepackage{graphicx}
\usepackage{subfigure}
\usepackage{booktabs}
\usepackage{todonotes}
\usepackage{amsmath,amsfonts,amsthm}
\usepackage{textcomp}
\usepackage{xcolor}
\usepackage{multirow}
\usepackage{enumitem}
\usepackage{url}
\usepackage[ruled, linesnumbered]{algorithm2e}
\usepackage{bm}
\usepackage{diagbox}
\newtheorem{myDef}{\textbf{Definition}}

\AtBeginDocument{%
  \providecommand\BibTeX{{%
    \normalfont B\kern-0.5em{\scshape i\kern-0.25em b}\kern-0.8em\TeX}}}

\def\numx#1e#2{{#1}\mathrm{e}{#2}}

\copyrightyear{2022}
\acmYear{2022}
\setcopyright{acmlicensed}\acmConference[CIKM '22]{Proceedings of the 31st ACM International Conference on Information and Knowledge Management}{October 17--21, 2022}{Atlanta, GA, USA}
\acmBooktitle{Proceedings of the 31st ACM International Conference on Information and Knowledge Management (CIKM '22), October 17--21, 2022, Atlanta, GA, USA}
\acmPrice{15.00}
\acmDOI{10.1145/3511808.3557120}
\acmISBN{978-1-4503-9236-5/22/10}

\begin{document}


\title{GIFT: Graph-guIded Feature Transfer for Cold-Start Video Click-Through Rate Prediction
}

\author{Yi Cao}
\authornote{The first two authors are of equal contribution and in no particular order.}
\email{dylan.cy@alibaba-inc.com}
\affiliation{
  \institution{Alibaba Group}
  \country{}
}
\author{Sihao Hu}
\authornotemark[1]
\email{husihao26@zju.edu.cn}
\affiliation{
  \institution{Zhejiang University}
  \country{}
}

\author{Yu Gong}
\email{gy910210@163.com}
\affiliation{
  \institution{Alibaba Group}
  \country{}
}

\author{Zhao Li}
\email{zhao_li@zju.edu.cn}
\affiliation{
  \institution{Zhejiang University}
  \country{}
}

\author{Yazheng Yang}
\email{yazheng\_yang@zju.edu.cn}
\affiliation{
  \institution{Zhejiang University}
  \country{}
}

\author{Qingwen Liu}
\email{xiangsheng.lqw@alibaba-inc.com}
\affiliation{
  \institution{Alibaba Group}
  \country{}
}

\author{Shouling Ji}
\authornote{Shouling Ji is the corresponding author.}
\email{sji@zju.edu.cn}
\affiliation{
  \institution{Zhejiang University}
  \country{}
}

\renewcommand{\shortauthors}{Yi Cao et al.}

\begin{abstract}
Short video has witnessed rapid growth in the past few years in e-commerce platforms like Taobao. To ensure the freshness of the content, platforms need to release a large number of new videos every day, making conventional click-through rate (CTR) prediction methods suffer from the item cold-start problem. 
In this paper, we propose GIFT, an efficient \textbf{G}raph-gu\textbf{I}ded \textbf{F}eature \textbf{T}ransfer system, to fully take advantages of the rich information of warmed-up videos to compensate for the cold-start ones. Specifically, we establish a heterogeneous graph that contains physical and semantic linkages to guide the feature transfer process from warmed-up video to cold-start videos. The physical linkages represent explicit relationships, while the semantic linkages measure the proximity of multi-modal representations of two videos. We elaborately design the feature transfer function to make aware of different types of transferred features (\textit{e.g.}, id representations and historical statistics) from different metapaths on the graph.
We conduct extensive experiments on a large real-world dataset, and the results show that our GIFT system outperforms SOTA methods significantly and brings a \textbf{6.82}\% lift on CTR in the homepage of Taobao App. 
\end{abstract}



\maketitle

\section{Introduction}




In e-commerce platforms like Taobao, a short video produced by an author to share a lifestyle or showcase an item, will be displayed in the homepage to attract customers' attention for their potential click and payment. In the past year, the short videos have witnessed rapid development in Taobao: the number of short videos has grown five-fold to over ten million, and the velocity of daily new video production has accelerated from tens of thousands to hundreds of thousands, which has greatly improved the richness and freshness of short videos, as well as the coverage for products.






\begin{figure*}[tbp!]
\centering
\includegraphics[width=17.2cm]{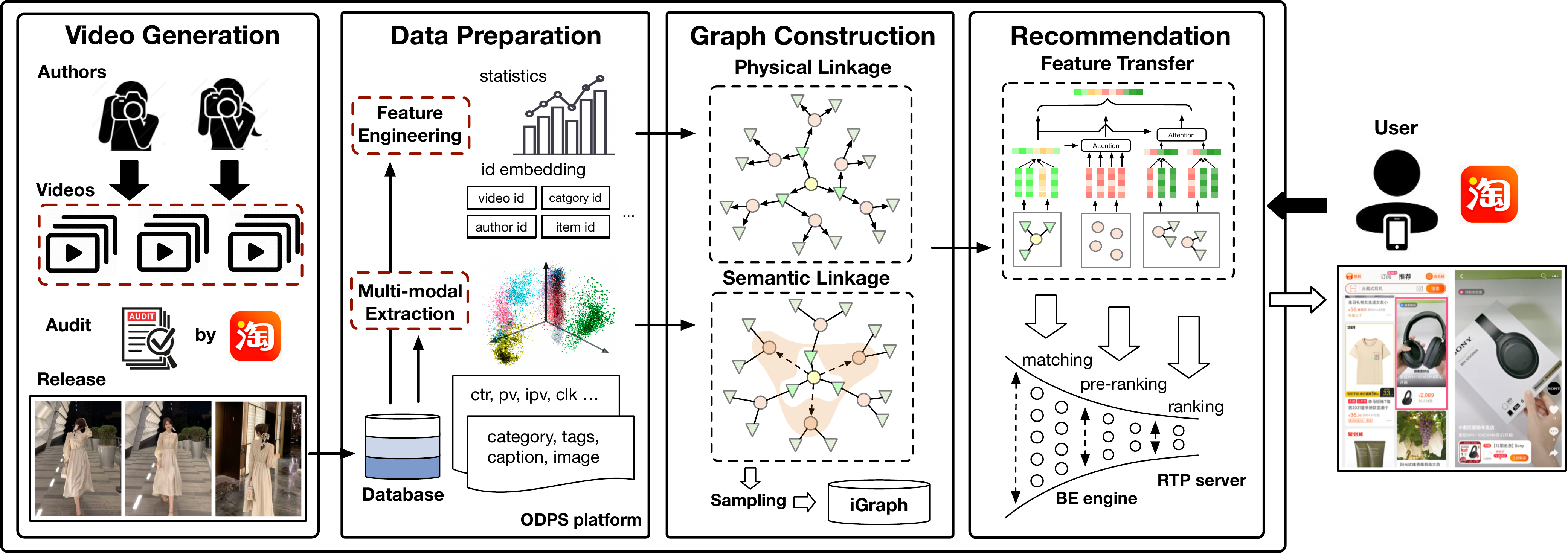}
\caption{Workflow of the whole cold-start video recommendation system in Taobao. The GIFT system works for graph construction and feature transfer stages.}
\label{fig:workFlowOfColdStartRS}
\vspace{-0.2cm}
\end{figure*} 

Behind this, a click-through-rate (CTR) model plays a vital role in accurately recommending appropriate videos for a target user. Unfortunately, the item cold-start problem\cite{gope2017survey,hu2022sequence} may occur when the model faces a large volume of new videos released every day. We unveil the underlying reasons for the existence of this problem: 1) Mainstream methods such as collaborative filtering algorithms\cite{koren2009matrix,su2009survey,yang2020large} require historical user-item interactions to calculate item co-occurrence relationship, leading to an unsatisfied result that new items that has no or rare user interactions can miss the opportunity to be recommended and remain cold all the time; 2) Deep learning methods\cite{YoutubeDNN,DIN,ZhaoliATNN,zhu2019addressing} learn a meaningful id embedding for each entity with at least 5-10 occurrences\cite{grbovic2018real} of that item, which is far beyond the average interaction times for the newly arrived videos; 3) An unbalanced percentage of cold-start videos in the total can cause the model to overweight the majority of non-cold-start videos, leading to poor performance on new videos.
Considerable efforts have been made to solve this problem, and most of them fall into the category of content-based methods\cite{van2000using,thorat2015survey,zhu2019addressing}, \textit{i.e.}, the performance lift comes from incorporating the content information like category, text caption, image, and video content representation. However, in the real-world industrial application, information mentioned above has long been incorporated to accurately portray an item, suggesting that content-based algorithms are no longer the silver bullet. Alternatively, the approach we seek is expected to capture some high-level information to represent a new video more precisely with the limited information. 

In this work, we present \textbf{G}raph-gu\textbf{I}ded \textbf{F}eature \textbf{T}ransfer (GIFT) system, with a straightforward idea to construct linkages to guide feature transfer from warmed-up videos to cold-start ones. The whole GIFT system is presented as in Figure~\ref{fig:workFlowOfColdStartRS}, which consists of the graph construction module and feature transfer model (GIFT network). For each video, it has the attribute like ``author'' and ``product'', which indicates its author and the product it showcase. Firstly, the graph construction module utilizes these attribute information to build edges between warmed-up videos and cold-start videos. Specifically, we introduce the concept of attribute node to represent a specific attribute value and video node to represent a unique video. We name an edge between a video node and an attribute node as a \textit{physical linkage} if the video has a specific attribute value. Based on this setting, a physical 
metapath ``V-A-V'' indicates two videos that have the same author, or showcase the same product.
By building physical metapaths, we can link more than 95\% of cold videos to at least one warm video, and this approach does not rely on any interaction logs limited for cold-start videos.

However, two concerns are raised about this construction way: 1) We cannot assure that all cold videos can link to enough warm videos ($\geq$5) for the effective transfer; 2) Physical metapath does not necessarily guarantee the most similarity in semantics, \textit{i.e.}, the shortest vicinity in the semantic space. To this end, we use a pre-trained multi-modal model to jointly encode the title and cover image of the video into a semantic space, where a vector represents a video. By simply applying $k$NN search, we can retrieve the most similar top-$k$ video nodes for each cold-start video node, which perfectly solve the above-mentioned problems. We name this relationship as \textit{semantic linkage}, and the semantic metapath ``V-V'' indicates two videos that are semantically related.



Guided by these two types of linkages, we can transfer features from warmed-up video nodes to the cold-start ones, not only the robust id representations, but also historical statistics like page view (PV) and click-through rate (CTR), \textit{etc}. For different metapaths, the linked video nodes exhibit different degree of relatedness, and different types of nodes visited along a metapath contain different information. To make the feature transfer process more precisely, we propose the GIFT network by elaborately designing the transfer function for different types of nodes visited along different metapaths.
Concretely, we apply the attention mechanism\cite{transformer} to extract the most informative feature from the video nodes and attribute nodes separately to ensure the awareness of different types of transferred features. 
Moreover, two types of physical metapaths and semantic metapath are separately modeled in the transfer process. We refer the above transfer strategy as the instance-level transfer. 


In practice, we first pretrain GIFT network on the whole set of videos to acquire the the robust id representations of warmed-up videos, and then fine-tune it on the logs of cold-start videos, to make it adapt to the target domain. We refer this transfer strategy as the model-level transfer.
By utilizing the above-mentioned instance- and model-based transfer strategies, GIFT achieves \textbf{1.25} AP gain of AUC over the base model, which is a huge boost for the CTR prediction task.
Additionally, the whole GIFT system has been deployed on Taobao since Sep. 21, 2020, serving for the cold-start video recommendation in the homepage of Taobao App. The gain of \textbf{6.82}\% on CTR metric further illustrates its effectiveness.

\noindent \textbf{Contributions.} To summarize, the contributions are as follows:
\begin{itemize}[leftmargin=10 pt]

  \item We propose physical and semantic linkage construction methods to guide feature transfer process, with very limited information used and no requirement for any user interaction logs.
  
  

  \item We elaborately design the GIFT network to force the model effectively extract and transfer id representation and statistical features to the cold-start videos. 

  \item We conduct extensive experiments on a massive real-world dataset and Taobao's online environment, suggesting that GIFT achieves a significant CTR improvement and satisfying time efficiency and scalability for the real-world application.

  
\end{itemize}

\begin{figure*}[tbp!]
\centering
\includegraphics[width=17.2cm]{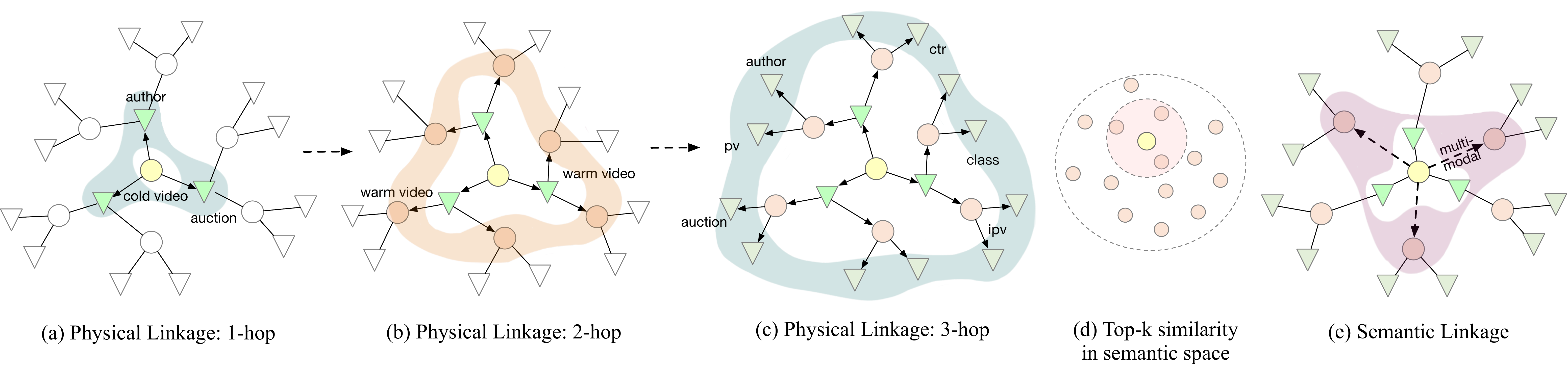}
\caption{Construction rules for two types of linkages. (a)-(c): physical linkage construction; (d): toy example for top-k similarity in the semantic representation space;(e): semantic linkage construction based on top-k similarity.}
\vspace{-0.2cm}
\label{fig:GraphConstruction}
\end{figure*}

\section{Preliminaries}

\subsection{Item Cold-start Problem}

We focus on addressing the item (video) cold-start problem, which is the problem that new items has no or rare prior events, like rating or click logs, making them miss the opportunity to be recommended and remain “cold” all the time. In our application, a \textit{new} or \textit{cold-start video} is defined as a short video released less than 3 days (inclusive) in Taobao. 



\vspace{-0.2cm}
\subsection{Workflow of Cold-start Video Recommendation System}
Figure~\ref{fig:workFlowOfColdStartRS} is the workflow of the whole cold-start video recommendation system, which consists of four parts: video generation, data preparation, graph construction, and recommendation. Our GIFT system works for the latter two parts. In video generation process, after produced by authors and audited by the platform, new videos can be added up to the video pool and recommended to users. Then the data preparation module calculates the statistics of videos periodically, as well as other features like multi-modal representations extracted based on videos' content.
Third, our GIFT system constructs a heterogeneous graph by building physical and semantic linkages between new and warmed-up videos. The computation graph for each cold video is pre-sampled and stored in Alibaba's iGraph database to support real-time inference.
Finally, in recommendation process, we use GIFT network to enhance the robustness of video feature to precisely predict the CTR for each item given the user. A typical industrial recommendation pipeline follows a multi-stage paradigm to trade off the accuracy and computation overhead (Section~\ref{sec:implementation}). In each stage, the model predicts CTR for each given item and then selects top ranked ones for the following stages. In theory, GIFT network works for all the stages. In practice, we implement and deploy it in the \textit{ranking} stage.


\section{Graph Construction}
\label{sec:graph_construct}
    
    Graph is a ubiquitous data structure to represent the relationships between entities in e-commerce\cite{ZhaoliBipartite,Turbo,MERIC}.
    In GIFT system, we construct a heterogeneous graph that consists of two physical metapath and one semantic metapath to guide the feature transfer process. In the graph, a video node uniquely identifies a video and an attribute node represents a specific attribute value.
    


\subsection{Physical Linkage}
\label{sec:physicalLinkage}

    The overall construction of physical linkage can be divided into three sub-steps: one-hop, two-hop, and three-hop linkage construction as illustrated in Figure~\ref{fig:GraphConstruction}(a)-(c), where the central round yellow node represents the target cold-start video. 

    \noindent \textbf{One-hop linkage:}
    As presented in Figure~\ref{fig:GraphConstruction}(a), the one-hop linkages are naturally set between a target video node (yellow round) and attribute nodes (green triangle), indicating that a video has certain attribute values, \textit{e.g.}, a video has an attribute of author id, category id, product id of the product it showcase, and other historical statistic features like average user stay time, page view (PV) and click-through rate (CTR) that calculated based on historical user behavior logs. It should be noted that the statistics of cold-start videos are unreliable. If we directly make use of these information without expanding the computation graph outward, it is a typical implementation of the content-based methods.

    
    \noindent \textbf{Two-hop Linkage:}
    As in Figure~\ref{fig:GraphConstruction}(b), the two-hop linkages expand to warmed-up video nodes (pink round nodes) from two certain types of attribute nodes, \textit{i.e.}, author id and product id.
    Empirically, videos photographed by the same author or showcasing the same product may have very similar content and style, thus more likely to be clicked by the same group of users and tend to have similar recommendation pattern.
    In our implementation, the video nodes connected by two-hop linkages are constraint as the warmed-up videos, which have been released for more than 3 days so that they are able to accumulate enough user behavior logs and the statistics are more reliable. By building the two-hop linkage we are able to link more than 95\% of cold videos to at least one warm video.
    

    
    \noindent \textbf{Three-hop Linkage:}
    If the model solely takes the two-hop computation graph as input, it can be regarded as a representation transfer model, where the transferred features are robust representations of the two-hop linked video nodes.
    As presented in Figure~\ref{fig:GraphConstruction}(c), we take a step outward by setting three-hop linkage from the outermost warmed-up video nodes to neighbored attribute nodes, just the same rule as the one-hop linkage. Three-hop linkage can link to attribute nodes that represent informative value of the warmed-up videos like historical statistics like page view (PV), average stay time, click-through rate (CTR), etc., which are exactly what cold-start video lacks and needs to be made up for.
    
    
    






\begin{figure*}[tbp]
\centering
\includegraphics[width=17.3cm]{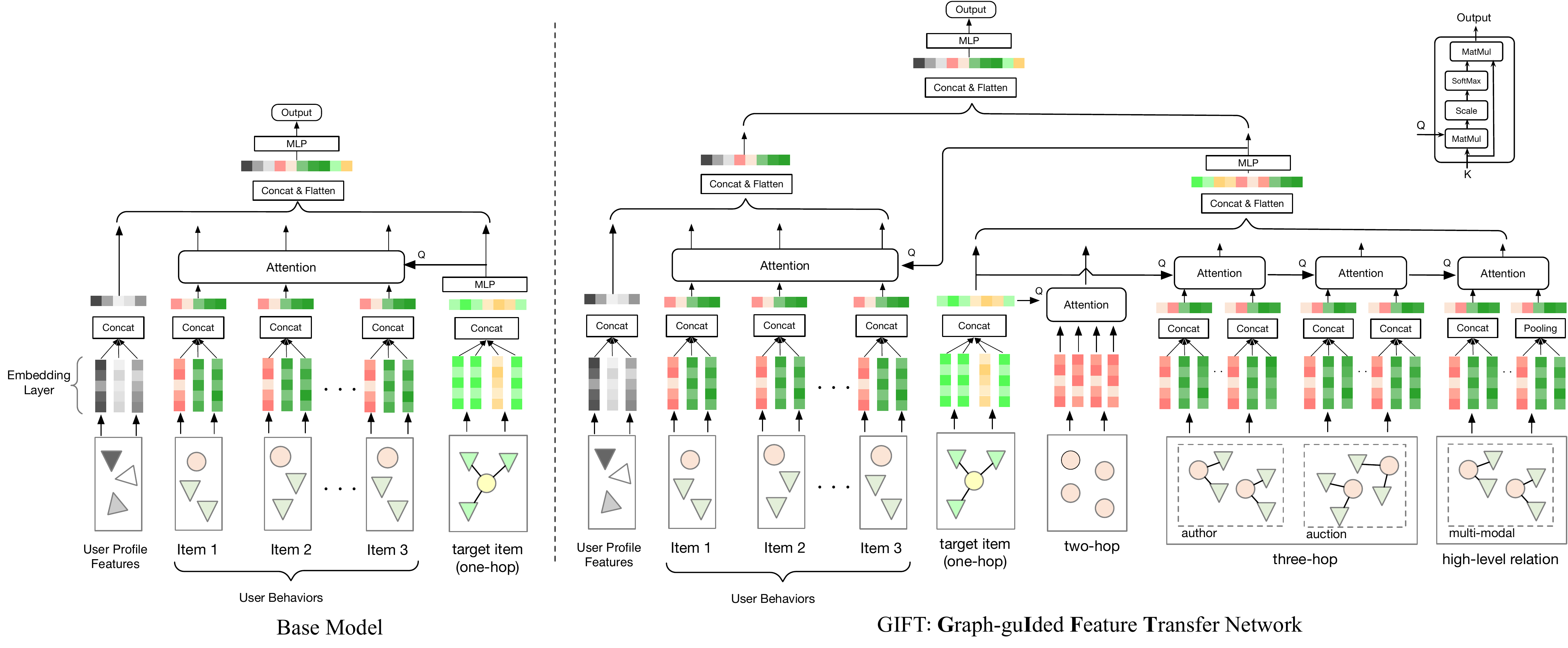}
\caption{Metapath-guided Graph Transfer Neural Network: Left is the user-side model and right is the item-side model.}
\vspace{-0.2cm}
\label{fig:GraphTransferModel}
\end{figure*} 

\subsection{Semantic Linkage}
\label{linkagesection}

The above method solely focuses on the linkages that physically exist, but there are still two concerns about this construction way: 1) We cannot assure that all cold-start videos can link to at least one warmed-up video; 2) Video nodes linked by physical linkages does not necessarily guarantee the shortest vicinity in the semantic space, \textit{i.e}, they may not the most similar videos in content or style. Therefore, we propose the \textbf{semantic linkage} construction way to solve these two problems further. 
By taking two factors into account, \textit{i.e.}, the title and cover image of the video, we use a pre-trained multimodal model lxmert\cite{tan2019lxmert} to encode these two types of information into a semantic space, where a vector represents a video as presented in Figure~\ref{fig:GraphConstruction}(d). Consequently, $k$NN search can be easily applied into this space to acquire $k$ videos that relate to the target video without any requirement of physical relationship. Moreover, the vicinity calculated through metrics like cosine distance ensures the top-$k$ similarity in the multimodal semantic space. By taking the semantic-neighbored videos as the source node, we further links the attribute nodes one-step outward, which represents the historical statistics of PV, CTR and so on, just the same rule as the three-hop physical linkage in Section~\ref{sec:physicalLinkage}. 






\section{CTR Prediction Model}
\label{sec:ranking}
Existing approaches\cite{YoutubeDNN,DIN,zhou2019deep,pi2019practice} can be categorized as user-side and item-side models. In this section, the proposed GIFT network focuses on the item (video)-side designing, because it addresses the item cold-start problem. We select DIN\cite{DIN} as our base user-side model, considering it is much more widely used in Taobao's applications than some newer models like DIEN\cite{zhou2019deep}, MIMN\cite{pi2019practice} due to its online efficiency and effectiveness.
However, it should be noted that GIFT does \textbf{not couple with} any user-side models and can be easily generalized to SOTA methods.

\subsection{Base Model: DIN}

DIN has been successfully applied in Taobao. It mainly proposes the target attention mechanism to adaptively learn the representation of user interests from historical behaviors \textit{w.r.t.} the target item, as shown in the left of Figure~\ref{fig:GraphTransferModel}. It consists of several parts:

\noindent \textbf{Embedding Layer:} The input of DIN in our scenario contains several categorical id features, \textit{e.g.}, video id, item id, etc. These id features cannot be directly input into the model because of their extremely high dimensionality. For instance, the number of item ids is about billions. Therefore, the widely-used embedding technique is adopted to embed the original sparse features into low-dimensional dense vectors, which significantly eases the model computing process. For user and items (both the target item and items in user behavior sequence), id embeddings are concatenated with its other continuous features together to generate the overall embedding as depicted in the left of Figure~\ref{fig:GraphTransferModel}.



\noindent \textbf{Target Attention Mechanism:} The original DIN proposes a local activation unit to calculate the representation vector of user interests by considering the relevance of historical behaviors \textit{w.r.t.} the candidate item, which means the representation vector varies over different candidate items. This local activation unit can be re-implemented under the framework of attention\cite{transformer} by considering the representation of target item $\boldsymbol{h}_t$ as the query and $\boldsymbol{H}_{u}$ as key and value, where $\boldsymbol{H}_{u} = \{\boldsymbol{h}_{1}, \boldsymbol{h}_{2}, \ldots,\boldsymbol{h}_{H}\}$ is the set of embedding vectors of items in the user behaviors with length of $H$. Specifically, we formalize it as follows:
\begin{equation}
\boldsymbol{h}_u =\text{TargetAttention}\left(\boldsymbol{h}_{t} \boldsymbol{W}^{Q}, \boldsymbol{H}_{u} \boldsymbol{W}^{K}, \boldsymbol{H}_{u} \boldsymbol{W}^{V}\right)
\end{equation}
\begin{equation}
\text{TargetAttention}(Q, K, V)=\operatorname{softmax}\left(\frac{Q K^{\top}}{\sqrt{d}}\right) V
\end{equation}
where the projections matrices $\boldsymbol{W}^{Q} \in \mathbb{R}^{d \times d}$, $\boldsymbol{W}^{K} \in \mathbb{R}^{d \times d}$, $\boldsymbol{W}^{V} \in \mathbb{R}^{d \times d}$ are learnable parameters and $d$ is the dimension of hidden space. $\boldsymbol{H}_{u}$ can be regarded as a matrix $\in \mathbb{R}^{d \times H}$, and $\boldsymbol{h}_t \in \mathbb{R}^{d}$ is the hidden representation of target item that passed through an multi-layer network. The temperature $\sqrt{d}$ is introduced to produce a softer attention distribution for avoiding extremely small gradients. 


\noindent \textbf{Loss:} The objective function used in DIN and GIFT is both the negative log-likelihood function defined as: 
\begin{equation}
L=-\frac{1}{N} \sum_{(x, y) \in \mathcal{D}}(y \log f(x)+(1-y) \log (1-f(x)))
\end{equation}
where $D$ is the training set, with $x$ as the input of the model and $y\in\{0,1\}$ as the ground-truth label, $f(x)$ is the output after the softmax layer that represents the predicted probability whether sample $x$ is clicked.

\subsection{GIFT: Graph-guided Feature Transfer}
\label{sec:gift}

The overall structure of the GIFT network is demonstrated in Figure~\ref{fig:GraphTransferModel}. 
Given the sampled computation graph for a target video node, two motivations are adopted in our model designing: 1) For different metapaths, the linked video nodes exhbit different degree of relatedness; 2) Different types of nodes visited along a metapath contain different information. To make the feature transfer process more precisely, we need to clearly distinguish different types of nodes visited along different metapaths. 


\noindent \textbf{Metapath}\cite{MERIC}: metapath is defined as a relation sequence to capture the specific structural relation between objects. As illustrated in Section~\ref{sec:graph_construct}, there are total three metapaths in the graph: two physical metapaths $\rho_a$ and $\rho_p$, where $\rho_a=v_{t} \rightarrow a(author) \rightarrow v \rightarrow a$, $\rho_p=v_{t} \rightarrow a(product) \rightarrow v \rightarrow a$, and one semantic metapath $\rho_s=v_{t} \rightarrow v \rightarrow a$.
Metapath has been widely used in graph mining community to constrain a meta structure in a heterogeneous graph that represents a specific semantic pattern. We further introduce the concept of metapath-guided neighbors to represent sets of nodes visited along the given metapath $\rho$ since we focus more on nodes to transfer the feature than the metapath itself.
·
\begin{myDef} \textbf{Metapath-guided Neighbors}\cite{MERIC}. Given a node $o$ and a metapath $\rho$ (start from $o$) in the graph, the metapath-guided neighbors is defined as the set of all visited objects when the object $o$ walks along the given metapath. In addition, we denote the $i$-th step neighbors of object $o$ as $\mathcal{N}_{\rho}^{(i)}(o)$. 
For example, given the metapath $\rho_a=v_{t} \rightarrow a(author) \rightarrow v \rightarrow a$, we can get metapath-guided neighbors as $\mathcal{N}_{\rho_a}^{(1)}\left(v_{t}\right)=\left\{a_{1}, a_{2}\right\}, \mathcal{N}_{\rho_a}^{(2)}\left(v_{t}\right)=\left\{v_{1}, v_{2}, v_{3}\right\} $. $\mathcal{N}_{\rho_a}^{(0)}(v_{t})$ is $v_{t}$ itself.
\end{myDef}
\label{def}

It is worth noting that a video node represents the video id of itself and the attribute node represents a certain attribute value like category id. Following this setting, we use set of nodes to describe features used in our model, \textit{e.g.}, $\{v_t\} \cup \mathcal{N}_{\rho_a}^{(1)}(v_t)$ represents the id and attributes of video $v_t$. 




\noindent \textbf{Metapath-guided Feature Transfer:} With the help of metapath-guided neighbor, it is clear to describe the following feature transfer stage. 
To begin with, we obtain a representation vector $\boldsymbol{h}_{t}$ by embedding layer to represent the original information of target video $v_t$, which contains both id embedding and statistical features. For a cold-start item, this information is not robust due to its very limited interaction logs. Firstly, we transfer the id representation of the warmed-up video nodes to $\boldsymbol{h}_{t}$. The id representation of a warmed-up video is calculated based on sufficient user behavior logs, thus containing abundant information to transfer. The neighbored video nodes come from $\mathcal{N}^{(2)}_{\rho_a}(v_t)$, $\mathcal{N}^{(2)}_{\rho_p}(v_t)$, and  $\mathcal{N}^{(1)}_{\rho_2}(v_t)$. The generation of id representation $\mathbb{E}^{(2)}$ can be formalized as: 
\begin{equation}
\mathbb{E}^{(2)} = \left \{\text{Embed}\left(v_{i}\right) | v_{i} \in \mathcal{N}^{(2)}_{\rho_a}(v_t) \cup \mathcal{N}^{(2)}_{\rho_s}(v_t) \cup \mathcal{N}^{(1)}_{\rho_p}(v_t) \right \}
\end{equation}
where $\mathbb{E}^{(2)}$ can be deemed as a concatenated matrix of a set of representation vectors, and $\text{Embed}(\cdot)$ denotes the embedding layer operation.
Then, instead of mean or sum pooling $\mathbb{E}^{(2)}$ into a vector, we adopt the idea of target attention as follows:
\begin{equation}
\begin{aligned}
\boldsymbol{h}^{(2)} &=\text {TargetAttention}\left(\boldsymbol{h}_{t}\boldsymbol{W}^{Q}_{2}, \mathbb{E}^{(2)}\boldsymbol{W}^{K}_{2}, \mathbb{E}^{(2)}\boldsymbol{W}^{V}_{2} \right)
\end{aligned}
\end{equation}
Target attention function takes $\boldsymbol{h}_{t}$ as the query vector, and thus pay more attention to nodes which have similar representation to $\boldsymbol{h}_{t}$ during pooling, making $\boldsymbol{h}^{(2)}$ contain more useful information related to the target video $v_t$, compared to mean or sum pooling operations.

Since $\boldsymbol{h}^{(2)}$ now only include id representation information transferred from the warmed-up videos and do not contain any statistical information, we take a further step to get statistical features involved. To transfer statistical features, similarly, we generate the set of video representation $\mathbb{E}^{(3)}$ as follows:
\begin{equation}
\mathbb{E}^{(3)}_{a}= \left \{\text{Embed}\left(\{v\}\cup {\mathcal{N}_{\rho_a}^{(1)}(v)}\right) | v \in \mathcal{N}^{(2)}_{\rho_a}(v_t) \right \}
\end{equation}
\begin{equation}
\mathbb{E}^{(3)}_{p}= \left \{\text{Embed}\left(\{v\}\cup {\mathcal{N}_{\rho_p}^{(1)}(v)}\right) | v \in \mathcal{N}^{(2)}_{\rho_p}(v_t) \right \}
\end{equation}
\begin{equation}
\mathbb{E}^{(3)}_{s}= \left \{\text{Embed}\left(\{v\}\cup {\mathcal{N}_{\rho_s}^{(1)}(v)}\right) | v \in \mathcal{N}^{(1)}_{\rho_s}(v_t) \right \}
\end{equation}
\noindent where $\mathcal{N}_{\rho_a}^{(2)}(v_t)$, $\mathcal{N}_{\rho_p}^{(2)}(v_t)$ refers to the set of video nodes that two step away from $v_t$ along the metapath $\rho_a$ and $\rho_p$, respectively, and $\mathcal{N}_{\rho_s}^{(1)}(v_t)$ refers to the set of video nodes that one step away from $v_t$ along the metapath $\rho_s$. $\mathbb{E}^{(3)}_{a}$, $\mathbb{E}^{(3)}_{p}$ and $\mathbb{E}^{(3)}_{s}$ are three corresponding feature matrices for transfer. We then apply target attention to refine the information of $\mathbb{E}^{(3)}_{a}$, $\mathbb{E}^{(3)}_{p}$ and $\mathbb{E}^{(3)}_{s}$ separately. The step of this transfer process can be formalized as follows:
\begin{equation}
\label{eq:atten2}
\begin{aligned}
\boldsymbol{h}^{(3)}_r &=\text{TargetAttention}\left(\boldsymbol{h}_{t}\boldsymbol{W}^{Q}_{3,r}, \mathbb{E}_r^{(3)}\boldsymbol{W}^{K}_{3,r}, \mathbb{E}_r^{(3)}\boldsymbol{W}^{V}_{3,r} \right), r \in \left\{a,p,s \right\}
\end{aligned}
\end{equation}
\begin{equation}
\label{eq_compute_h3}
\boldsymbol{h}^{(3)} = \boldsymbol{h}^{(3)}_{a} \oplus \boldsymbol{h}^{(3)}_{p} \oplus \boldsymbol{h}^{(3)}_{s}
\end{equation}
It is noted that target attention are computed separately on $\mathbb{E}^{(3)}_{r}$ from metapath $\rho_r$, as illustrated in Figure~\ref{fig:GraphTransferModel}. This separate modeling strategy ensure the feature transfer only occurs within nodes that share a certain type of characteristic to prevent it from vanishing by mixing all the nodes together.
For example, videos that showcase the same product tend to share more similar content, whereas videos produced by the same author tend to have similar photographing styles. 
Finally, we concatenate $\boldsymbol{h}^{(3)}_a$, $\boldsymbol{h}^{(3)}_p$ and $\boldsymbol{h}^{(3)}_s$ into one vector to obtain the final $\boldsymbol{h}^{(3)}$ in Eq.~\ref{eq_compute_h3}, where $\oplus$ denotes the concatenate operator.

Given $v_t$'s transferred embedding features $\boldsymbol{h}^{(2)}$, $\boldsymbol{h}^{(3)}$ encoding different types of information from the warmed-up videos, we concatenate them with $\boldsymbol{h}_{t}$ and pass them through a MLP-layer to generate the final representation for $v_t$ as follows:
\begin{equation}
\boldsymbol{h}_{t}^{'} = \text{MLP}(\boldsymbol{h}_{t} \oplus \boldsymbol{h}^{(2)} \oplus \boldsymbol{h}^{(3)})
\end{equation}
Then we take $\boldsymbol{h}_{t}^{'}$ as the query vector and feed it into DIN. The remaining settings of the network structure are just the same as the base model. It is noted that due to the robustness of $\boldsymbol{h}_t'$, the efficiency of target attention is enhanced either. 






\subsection{Training Strategy}
\label{sec:training_strategy}

\noindent \textbf{Pre-training:} GIFT network serves for the short video newly launched on the Taobao, yet in order to acquire an effective model, we first pre-train the model on the whole set of videos, because it is needed to train robust representations for the warmed-up videos to facilitate the subsequent feature transfer. Moreover, because of the unbalanced distribution of the new and old video, user behavior logs collected from the new video (around 1/10 daily) are insufficient to train a model with a good generalization performance. 

\noindent \textbf{Fine-tuning:} However, domain bias between new and old videos not only exists in terms of data volume but also in the recommendation pattern: model trained on the logs of old videos more rely on the id embedding or historical statistics compared to the model trained on the new videos, which should assign more weights on the content information to well generalize to the cold-start videos. Therefore, we fine-tune our pre-trained model on the logs of cold-start videos collected over the past 1 month to obtain sufficient training logs. By using this adaptation strategy, our model boosts the offline performance by 0.86\% in terms of AUC (Table~\ref{tab:comparison}).


\section{System Implementation}
\label{sec:implementation}

\begin{figure}[tbp]
\centering
\includegraphics[width=4.6cm]{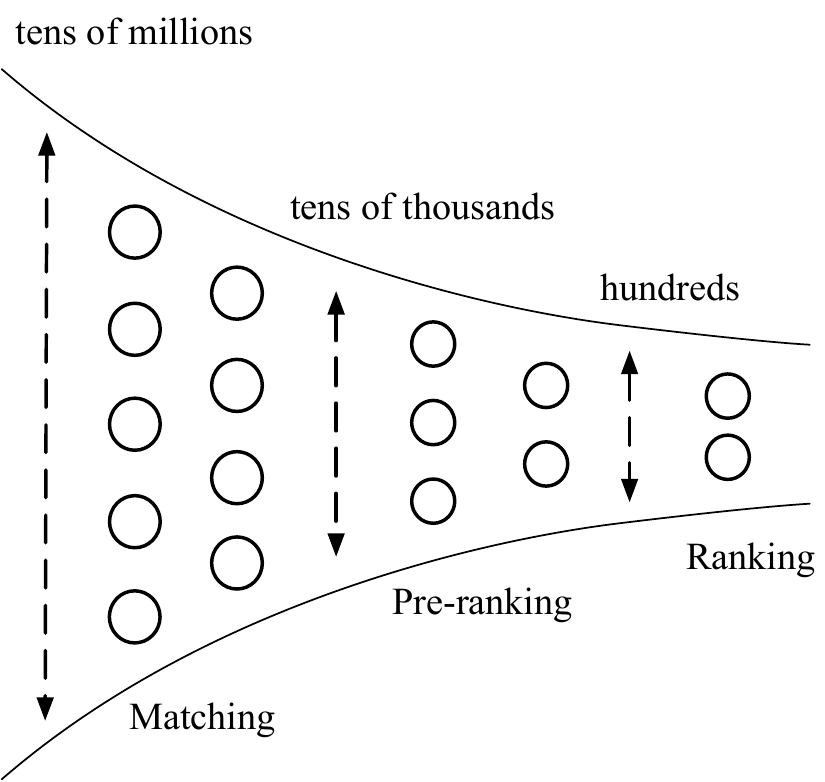}
\caption{The cascade architecture of industrial recommendation system.}
\label{fig:cascadeArchitecture}
\end{figure}

In industrial e-commerce scenarios, 
recommendation systems usually follow a multi-stage cascade architecture to trade off the accuracy and efficiency, \textit{i.e.}, first to use simpler models to select a candidate set from a large size of item pool and then use more complex models to filter the candidate set further.
As depicted in Figure~\ref{fig:cascadeArchitecture}, the whole process can be divided into three stages: \textit{matching}, \textit{pre-ranking}, and \textit{ranking}. 
1) In the matching stage, we adopt Swing\cite{yang2020large} to calculate similarity score for item-to-item retrieval, and implement MIND\cite{MIND} for user-to-item vector retrieval; 2) Pre-ranking can be seen as a simplified version of the ranking phase, considering the computation cost challenge of online serving with a larger size of the candidate set to be calculated (around tens of thousands). We implement the DNN\cite{YoutubeDNN} model for the pre-ranking stage in our pipeline; 3) Ranking stage plays a vital role in the whole cascade architecture. Since it tackles only hundreds of items, the model can take into account much more fine-grained features like cross-feature and adopt more intricate model architecture, like the target attention. In our implementation, the GIFT system works for the ranking stage and can be easily generalized to the matching and pre-ranking stages because it only involves calculation on the target item side. (interactive calculation between user- and item-side like target attention can not be supported by matching and pre-ranking stage because of the restriction on response time).

\begin{figure}[tbp]
\centering
\includegraphics[width=8.2cm]{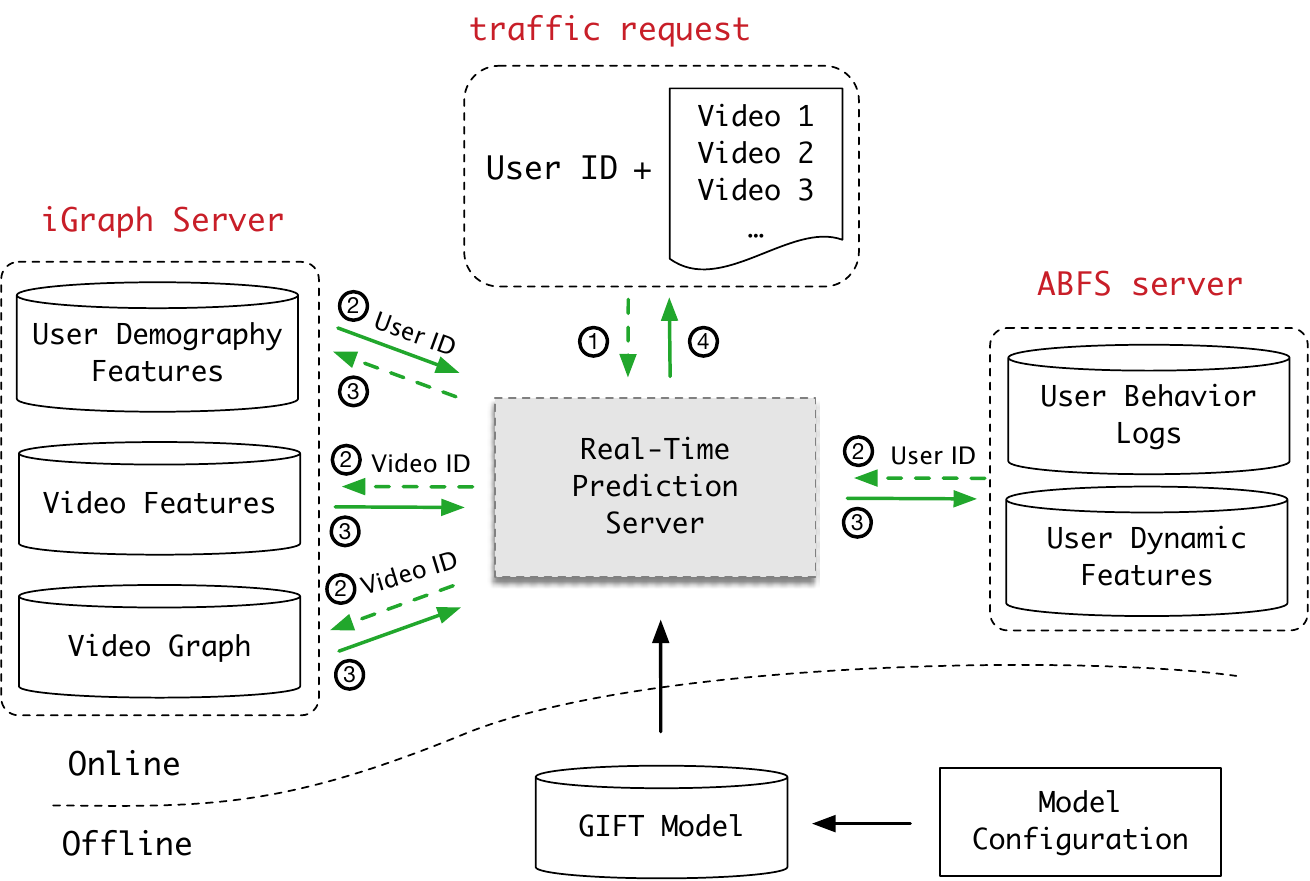}
\caption{The overall framework of GIFT system deployed in Taobao.}
\label{fig:framework}
\end{figure}  

Figure~\ref{fig:framework} gives a brief illustration of the GIFT system that has been deployed in the homepage of Taobao App, which consists of four key components: iGraph server, ABFS server, real-time prediction server and model management module. iGraph is an in-memory graph database developed by Alibaba Group, providing storage, real-time query, and update services for large-scale graph data; ABFS is a uniform feature service built by Alibaba Group, which can provide real-time user behavior logs with very low latency (tens of milliseconds) and generate dynamic user features like the length of stay time on a video.
Specifically, when a client makes a prediction request with the user id and the candidate videos selected by matching and pre-ranking models, the real-time prediction server queries the user demography features and video features stored in the iGraph server and the neighbored video nodes of the target video. Simultaneously, the real-time prediction server asks the ABFS server to return the clicked video sequence and the dynamic user feature. Taking above as the input, the prediction server makes a real-time prediction and returns to the client the predicted CTR for each candidate item. The execution order is marked by the number.

In our implementation, the user demography feature and video feature are relatively stable and thus can be updated in a low frequency like a daily basis. To keep the new videos fresh, we construct the video graph at a higher frequency of updates, \textit{e.g.}, hourly basis. It should be noted that the neighbored video nodes are sampled offline and pre-stored as a sequence for each cold-start video in iGraph server. To this end, there is no need to consider the overhead of graph sampling in online inference. As the CTR model in the ranking stage, GIFT is trained incrementally in a daily basis by the model management module. It is also very helpful to adopt online learning to improve the performance of new video recommendation in practice.

\begin{table*}[tbp]
\small
\centering
\caption{Statistics of Taobao Dataset}
\begin{tabular}{cc|ccc|cc|cc}
\toprule
\multicolumn{2}{c|}{\textbf{Dataset}}                       & \textbf{\# Users} & \textbf{\# Items} & \textbf{\# Samples} & \textbf{Edge Type} & \textbf{\# Edges} & \textbf{Path Type} & \textbf{\# Paths}  \\ 
\midrule 
\multicolumn{1}{l|}{\multirow{3}{*}{Taobao}} 
& full & $4.98 \times 10^{7}$  & $2.2 \times 10^{7}$  & $5.78 \times 10^{8}$  & V-A &  $2.2 \times 10^{7}$ & V-A-V& $1.9 \times 10^{8}$ \\ \multicolumn{1}{l|}{}                        
& cold & $3.0 \times 10^{7}$  & $4.8 \times 10^{5}$   & $1.38 \times 10^{8}$  & V-P  & $2.1 \times 10^{7}$ &V-P-V& $5.7\times 10^{7}$   \\ \multicolumn{1}{l|}{}                        
& test & $2.1 \times 10^{6}$  & $1.2 \times 10^{5}$ & $9.4 \times 10^{6}$ & V-V &  $2.8 \times 10^{8}$ & V-V & $2.8 \times 10^{8}$ \\ 
\bottomrule

\end{tabular}
\label{tab:datase_statistic}
\end{table*}

\begin{table}
\centering
\caption{Performance Comparison}
\begin{tabular}{lp{0.5m}p{0.55cm}cc}  
\toprule
\multicolumn{2}{c}{\diagbox{\textbf{Methods}}{\textbf{Metrics}}}& \multicolumn{1}{|c}{\textbf{AUC}} & \multicolumn{1}{c}{\textbf{RelaImpr}}   \\ 
\midrule
\multirow{3}{*}{\begin{tabular}[c]{@{}l@{}}Handcrafted\\Features\end{tabular}} &\multicolumn{1}{|c|}{LR} & 0.7218 & -13.63\% \\ 
                                                                                &\multicolumn{1}{|c|}{SVM} & 0.7339 & -8.92\%\\ 
                                                                                &\multicolumn{1}{|c|}{GBDT} & 0.7377 & -7.44\% \\ 
                                                                                
\midrule
\multirow{4}{*}{\begin{tabular}[c]{@{}l@{}}DNNs \end{tabular}} &\multicolumn{1}{|c|}{DNN}   & 0.7423 & -5.65\% \\ 
                                            
                                            &\multicolumn{1}{|c|}{Wide\&Deep} & 0.7465 &  -4.01\%  \\ 
                                                               &\multicolumn{1}{|c|}{DeepFM} & 0.7508 & -2.33\% \\ 
                                                          
                                            &\multicolumn{1}{|c|}{DIN} & \underline{0.7568} & \underline{0.00\%} \\     
\midrule
\multirow{2}{*}{\begin{tabular}[c]{@{}l@{}}Cold-Start \\ Methods \end{tabular}} &\multicolumn{1}{|c|}{DropOutNet} & 0.7573 & 0.19\% \\ 
                                        
                                        &\multicolumn{1}{|c|}{ACCM} & 0.7550 & -0.70\% \\ 
                                        
\midrule
\multirow{2}{*}{\begin{tabular}[c]{@{}l@{}}Ours \end{tabular}}
&\multicolumn{1}{|c|}{\textbf{GIFT}}  & \textbf{0.7670} & \textbf{3.97\%} \\
&\multicolumn{1}{|c|}{\textbf{GIFT$^{1}$}} & \textbf{0.7693} &  \textbf{4.87\%} \\ 
\bottomrule
\end{tabular}
\label{tab:comparison}
\end{table}

\section{Experiments}

    
    
    


\subsection{Dataset and Competitors}


    
    \indent \textbf{Taobao Dataset:}
    Taobao dataset includes hundreds of millions of user interaction logs with short videos in the homepage of Tabao App.
    The dataset consists of two parts: D$_{full}$ and D$_{cold}$, where D$_{full}$ is uniformly sampled from 15-day user interaction logs of all videos, and D$_{cold}$ is a subset of D$_{full}$ that only includes user interaction logs on new videos that launched less than 3 days. Both D$_{full}$ and D$_{cold}$ are used for the training phase. We further sample the testing set D$_{test}$ from the following one-day logs after D$_{full}$ and D$_{cold}$ are collected, which also only includes logs of cold videos. It is worth noting that the collection of this dataset strictly simulates the online environment.

\noindent \textbf{Competitors:} We compare proposed GIFT network with three types of baselines: the first type is conventional machine-learning methods based on handcrafted features, like Logistic Regression (LR), Support vector machine (SVM), Gradient Boosting Decision Tree (GBDT). The second type is DNN-based methods, including:
\begin{itemize}[leftmargin=8pt]
\item DNN\cite{YoutubeDNN}: a video recommendation approach proposed by Youtube and has been widely adopted in industrial application.


\item Wide\&Deep\cite{cheng2016wide}: a method proposed by Google that adopted a wide model to tackle manually designed cross-product features and a deep model to capture high-order feature interactions.

\item DIN\cite{DIN}: our base model as described in Section~\ref{sec:ranking}, which uses target attention to learn the representation of user interests from historical behaviors \textit{w.r.t.} the target item.

\end{itemize}

As for the third type of baselines, we implement two cold-start approaches: 
\begin{itemize}[leftmargin=8pt]


\item DropOutNet \cite{volkovs2017dropoutnet}: a method that applies dropout technique to make DNNs generalize to the missing input that is common for the cold-start items. DIN is implemented as the base model of DropOutNet.

\item ACCM\cite{shi2018attention}: a hybrid model that attentively integrates id embedding and content information into one vector to adapt to both the warm and cold items. For the fairness of comparison, we re-implement its base model as DIN.
\end{itemize} 

We do not put any pure content-based models into the competitors because the existing feature systems in Taobao has included abundant content information long before.

\noindent \textbf{Implementation Details:} We implement the DNN part of DeepFM, Wide\&Deep, DNN, DIN and GIFT just the same architecture, \textit{i.e.}, a three-layer MLP with 512, 256 and 128 hidden units. For all attention layers in above models, we set the number of hidden units to 128. Adagrad optimizer is adopted in all the methods, the learning rate of 1e-4 is set. Batch size of 512 among 256, 512, 1024 gives the best result for all DNNs. For other baselines, the grid search strategy is applied to find the optimal hyper-parameters.
For a certain type of physical linkage, we sample top-20 neighbors according to the ascending order of time interval between them and the target item; For semantic linkage, we sample top-20 neighbors according to the descending order of cosine similarity. The default value of 20 is chosen based on the observation that there are only marginal improvements when we continually increase this value.






\subsection{Performance Comparison}



In the field of CTR prediction, AUC value is widely used as a metric of goodness of order by ranking all the items with predicted CTR\cite{DIN,zhou2019deep}. We further adopt the idea of RelaImpr\cite{DIN} to measure the relative improvement over models. Since for a random guesser, the value of AUC is 0.5, the RelaImpr is defined as below:
\begin{equation}
    \text { RelaImpr }=\left(\frac{\text { AUC(measured model })-0.5}{\text { AUC(base model) }-0.5}-1\right) \times 100 \%
\end{equation}

Table~\ref{tab:comparison} shows the results of ten methods on the testing dataset. 
Obviously, all the DNNs beat non-deep learning models significantly, which demonstrates the power of deep learning in recommendation field. For DNN-based methods, Wide\&Deep with elaborately designed "wide" structure outperforms DNN, and DeepFM performs better than Wide\&Deep further; DIN stands out significantly among all the DNN-based methods, especially on Taobao Dataset with rich user behaviors.
We also observe that two cold-start methods do not work well on Taobao dataset: DropOutNet outperforms the base model with only 0.19\% of RelaImpr, ACCM performs even worse than DIN with -0.70\% of RelaImpr.
Instead, GIFT achieves superior performance compared with all the competitors. The performance boost mainly comes from the graph construction and feature transfer mechanism, which can be observed by comparing GIFT and its base model DIN (\textbf{0.01} absolute AUC gain and \textbf{3.97}\% of RelaImpr on Taobao dataset), and shows it does bring the information that cold-start videos covet most. We also compare GIFT to the GIFT$^1$ that fine-tunes on the $D_{cold}$ to reduce the domain bias, and an \textbf{0.86}\% RelaImpr is observed on the testing set. 
In e-commercial recommendation systems with hundreds of millions of traffic, 0.01 absolute AUC gain is significant enough and worthy of model deployment.

\subsection{Ablation Study}


\noindent \textbf{Feature-level Ablation:} It is easy to wonder what kind of transferred features play the most crucial role in feature transfer. The transferred features can be divided into two categories as demonstrated in Table~\ref{tab:ablation_feature}. 
The representation features of warmed-up video are dense vectors generated by the embedding layer and compensate for the inadequate training of representation feature of new videos.
Statistical features represent the historical performance of the video in Taobao, which are completely absent from new videos.
In the experiment, we remove these two types of features in turn and evaluate the performance of ablated transferred features. 


The results are presented in Table~\ref{tab:ablation_graph}, where we use (-) to denote the removed part. Specifically, \textbf{R}(-) means removing the representation feature from the transferred feature, and $\textbf{S}$(-) denotes removing the statistical features. From the table, we can observe that AUC drops 2.63\% of RelaImpr, which means the performance boost of GIFT owes much to the id representation. We conjure the reason is that the representation features are not only important for new videos, but also essential for attention calculation, \textit{i.e.}, it is difficult to calculate the accurate attention scores just based on the fixed numeric statistical features.
We also observe that $\mathbf{S}$(-) drops 1.27\% of RelaImpr, which suggests that historical statistics of the warmed-up video are also a very important supplement to the original feature of new videos.

\begin{table}[tbp]
\small
\centering
\caption{Description of Transferred Video Feature}
\begin{tabular}{l|l|c|c}
\toprule
\textbf{Granularity} & \textbf{Feature Name}   & \multicolumn{1}{l|}{\textbf{Dim}} & \textbf{Type} \\ 
\midrule 
\multirow{6}{*}{\begin{tabular}[c]{@{}l@{}}Representation\\ Feature \\ \end{tabular}} & video\_id                & 32                       & one-hot   \\
                                                                                  & item\_id             & 32                       & one-hot   \\
                                                                                  & author\_id               & 16                       & one-hot   \\
                                                                                  & category\_id                 & 16                       & one-hot   \\
                                                                                  & title\_token\_ids        & 16                       & multi-hot \\
                                                                                  & \multicolumn{1}{c|}{...} & ...                      & ...       \\ \midrule
\multirow{4}{*}{\begin{tabular}[c]{@{}l@{}}Statistical\\ Features\\  \end{tabular}}   
& ctr\_15d & 1  & numeric \\
& pv\_cnt\_15d & 1 & numeric \\
& ipv\_cnt\_15d  & 1   & numeric \\
& clk\_cnt\_15d  & 1  & numeric \\
& \multicolumn{1}{c|}{...} & ..                       & ..        \\ \bottomrule
\end{tabular}
\label{tab:ablation_feature}
\end{table}

\noindent \textbf{Model-level Ablation:} We remove transferred features generated by different phases of model in turn to evaluate the performance of ablated models.
Results are presented in Table~\ref{tab:ablation_graph}, where $\boldsymbol{h}^{(2)}$(-) means removing the transferred embedding $\boldsymbol{h}^{(2)}$ but keeping $\boldsymbol{h}^{(3)}$ and $\boldsymbol{h}^{(3)}$(-) vice versa. 
From the table, we can see that the performance drops when either $\boldsymbol{h}^{(2)}$ or $\boldsymbol{h}^{(3)}$ is removed (-1.69\% and -2.85\% of RelaImpr). Removing both of them impacts the performance more significantly, suggesting that the two
transferred embeddings work well together to generate more expressive video representations. 
Another observation is that the $\boldsymbol{h}^{(3)}$ contribute more than $\boldsymbol{h}^{(2)}$ to the final performance lift. This is reasonable because $\boldsymbol{h}^{(2)}$ only contains the \textit{video id} representation of neighbored videos, but $\boldsymbol{h}^{(3)}$ includes not only other id information but also the statistical feature the cold video lacks.

\noindent \textbf{Graph-level Ablation:} We study the contribution of each type of linkages by masking a specific type of linkages in turn. The results are presented in Table~\ref{tab:ablation_graph}, where $\mathcal{P}(author)$(-) means removing the physical linkages built by the same author relation and $\mathcal{P}(product)$(-) means removing edges built by the same product relation. $\mathcal{P}$(-) means removing both of these two types of physical linkages, and $\mathcal{S}$(-) means removing the type of semantic linkages. As we can see, the performance of $\mathcal{P}(author)$(-) exceeds $\mathcal{P}(
product)$(-) with 0.42\% of RelaImpr, which suggests $\mathcal{P}(product)$ is a stronger relationship compared to $\mathcal{P}(author)$ for feature transfer. We also observe that semantic linkages contribute a lot to the performance lift (the AUC drop reaches -1.80\% of RelaImpr), which suggests that physical and semantic linkages can compensate for each other to conduct more feature transfer.

\begin{table}[tbp]
\small
  \centering
  \caption{Ablation Study on Feature-, Graph- and Model-level}
  \begin{tabular}{c|c|c c c c c}
  \toprule
  \textbf{Category} &\textbf{Operator} &\textbf{AUC} & \textbf{RelaImpr}\\ 
  \midrule
  
  \multirow{1}{*}{\begin{tabular}[c]{@{}l@{}} - \end{tabular}} 
  & Base Model & $0.7568$ & -$3.82\%$ \\ 
  \midrule
  \multirow{2}{*}{\begin{tabular}[c]{@{}l@{}} Feature \end{tabular}} 
  &$\mathbf{R}$(-) & $0.7600$ & -$2.62\%$ \\
  &$\mathbf{S}$(-) & $0.7636$ & -$1.27\%$ \\
  \midrule
  \multirow{5}{*}{\begin{tabular}[c]{@{}l@{}} Graph \end{tabular}} &$\mathcal{P}$(-) & $0.7603$ & -$2.51\%$ \\
  &$\mathcal{P}(author)$(-) & $0.7631$ & -$1.46\%$ \\
  &$\mathcal{P}(product)$(-) & $0.7620$ & -$1.87\%$ \\
  &$\mathcal{S}$(-) & $0.7622$ & -$1.80\%$ \\
  \midrule
  \multirow{2}{*}{\begin{tabular}[c]{@{}l@{}} Model \end{tabular}} 
  &$\boldsymbol{h}^{(2)}$(-) & $0.7635$ & -$1.31\%$ \\
  &$\boldsymbol{h}^{(3)}$(-) & $0.7594$ & -$2.85\%$ \\
  \midrule
  \multirow{1}{*}{\begin{tabular}[c]{@{}l@{}} - \end{tabular}} 
  & GIFT & $0.7670$ & $0.00\%$ \\
  \bottomrule 
  \end{tabular}
  \label{tab:ablation_graph}
\end{table}



\subsection{Applying GIFT to Other Models}

\begin{figure}[tbp]
\centering
\includegraphics[width=7cm]{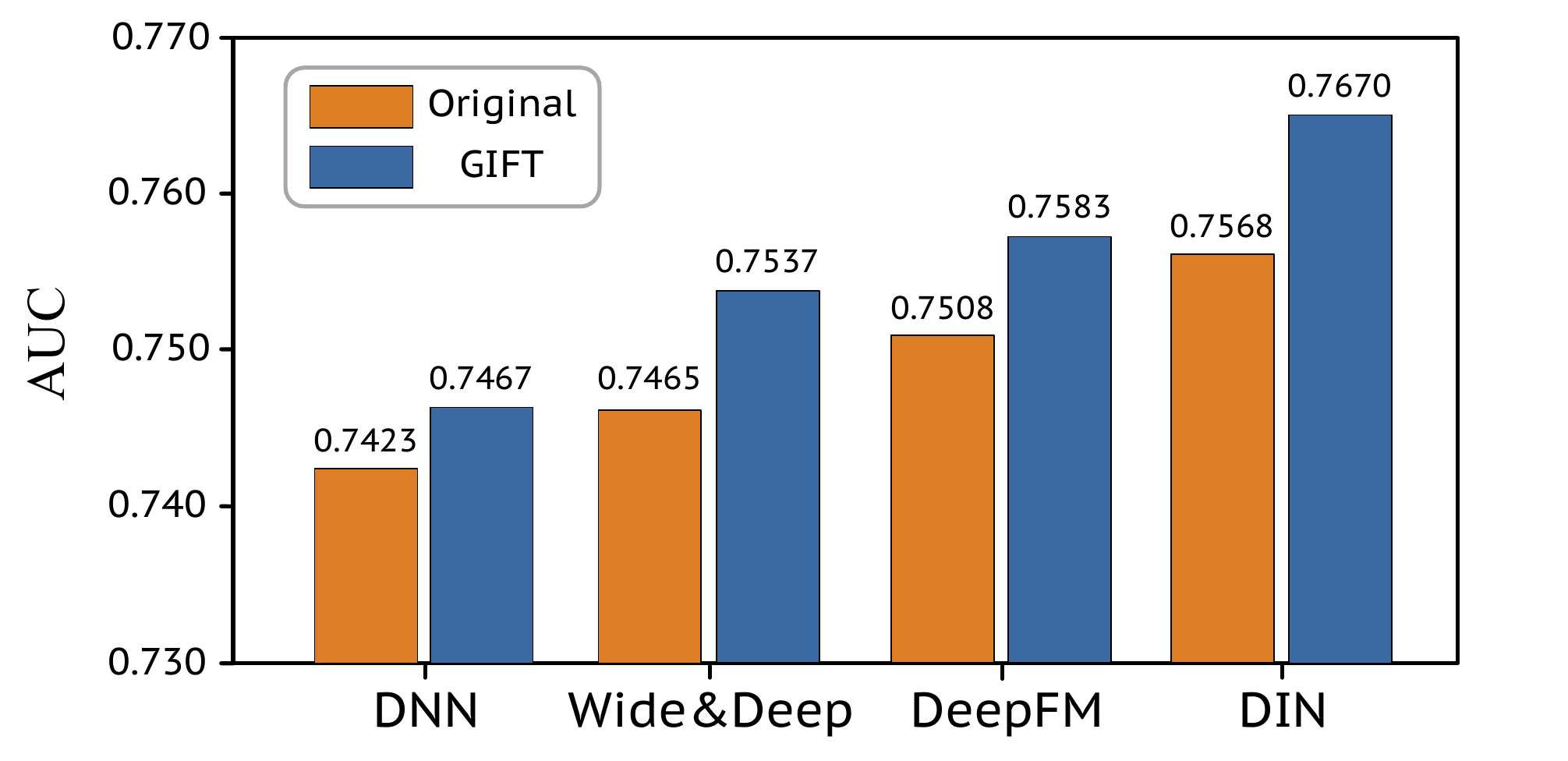}
\caption{Performance (AUC) comparison of different models enhanced by our GIFT method on Taobao dataset.}
\label{fig:GIFTBoost}
\end{figure}

GIFT can be generally apply to various models. To prove this, we conduct an experiment to examine whether it can bring improvements to other models. Specifically, we equip it for DNN-based models like DNN, Wide\&Deep, DeepFM, DIN on Taobao dataset. 
The results of AUC on Taobao dataset are shown in Figure~\ref{fig:GIFTBoost}. First, after applying with GIFT, all the baselines achieve better performance. This shows that item-side graph feature transfer can be easily applied to improve their performance on the cold-start CTR prediction task. Second, the performance lift of DIN ($3.97\%$ of RelaImpr) exceeds other baselines. This is because GIFT can enhance the robustness of video embedding, which improve the effectiveness of target attention by calculating more accurate attention scores between the target video and videos user have clicked.







\subsection{Study of Online A/B Test}
To verify the effectiveness of the proposed method in a real-world scenario, we implement and deploy GIFT for the cold-start video recommendation of the homepage of Taobao App. The system implementation is described in Section~\ref{sec:implementation}, and the baseline system adopts the DIN in the ranking stage with all the other conditions remain the same. 
We conducted the strict online A/B test on the live application spanning from 
Sep. 21, 2020, to Sep. 27, 2020 and involves hundreds of millions of users per day. Online evaluation shows that GIFT has achieved \textbf{6.82\%} lift on CTR metric (from 4.180\% to 4.465\%) with the 20\% of overall traffic on the test bucket and 80\% on the baseline bucket, which can bring huge economic benefits to the e-commerce platform.


\subsection{Study of Online Response Time}

We evaluate GIFT's efficiency in the real-world application by comparing the time cost between the baseline system and the system equipped with GIFT. The response time (in milliseconds) with thousands of queries per second during Sep. 23, 2020 is presented in Figure~\ref{fig:onlineRT}, where the gray and green lines represent the response time of the baseline system and our system. Empirical evidence shows that the response time of our system is only about 4 ms more than that of the baseline system on average, which is brought by the extra graph data retrieving and computational overhead boost of inference, yet the whole response time is still far below the maximum restriction of 300 ms. Since the neighbored graph information for each cold-start video is sampled offline and pre-stored in iGraph server, and the queries of graph data and features are asynchronous, the lift in response time brought by graph retrieving is negligible.


\subsection{Study of Scalability} 

\begin{figure}[tbp]
  
  \begin{minipage}{0.45\textwidth}
  \centering
  \includegraphics[width=6.5cm]{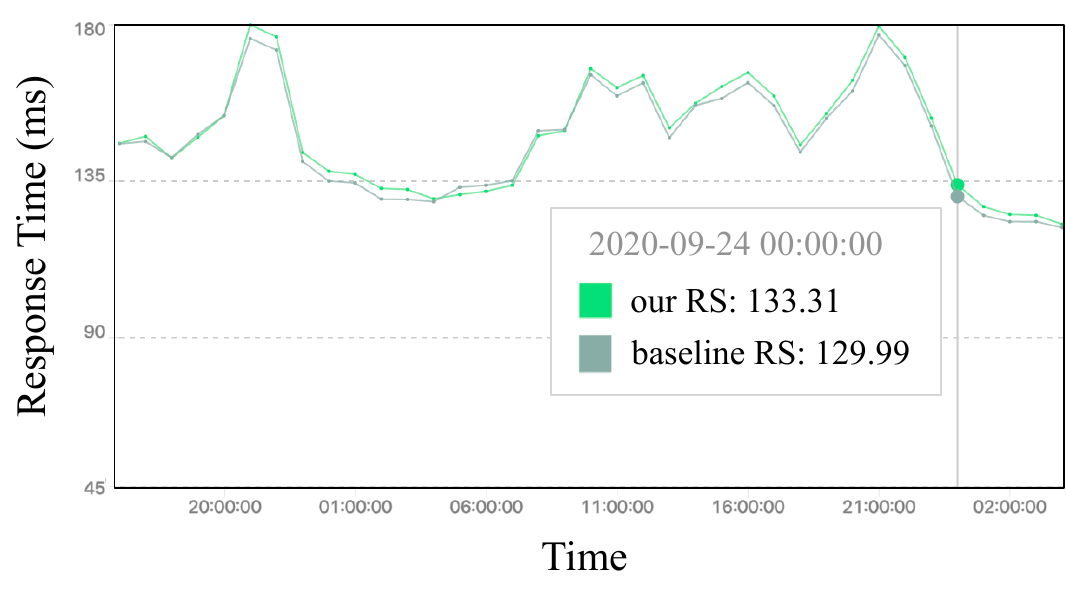}
  \caption{Comparison of Online Response Time}
  \label{fig:onlineRT}
  \end{minipage}
  \begin{minipage}{0.48\textwidth}
  \centering
  \includegraphics[width=8.7cm]{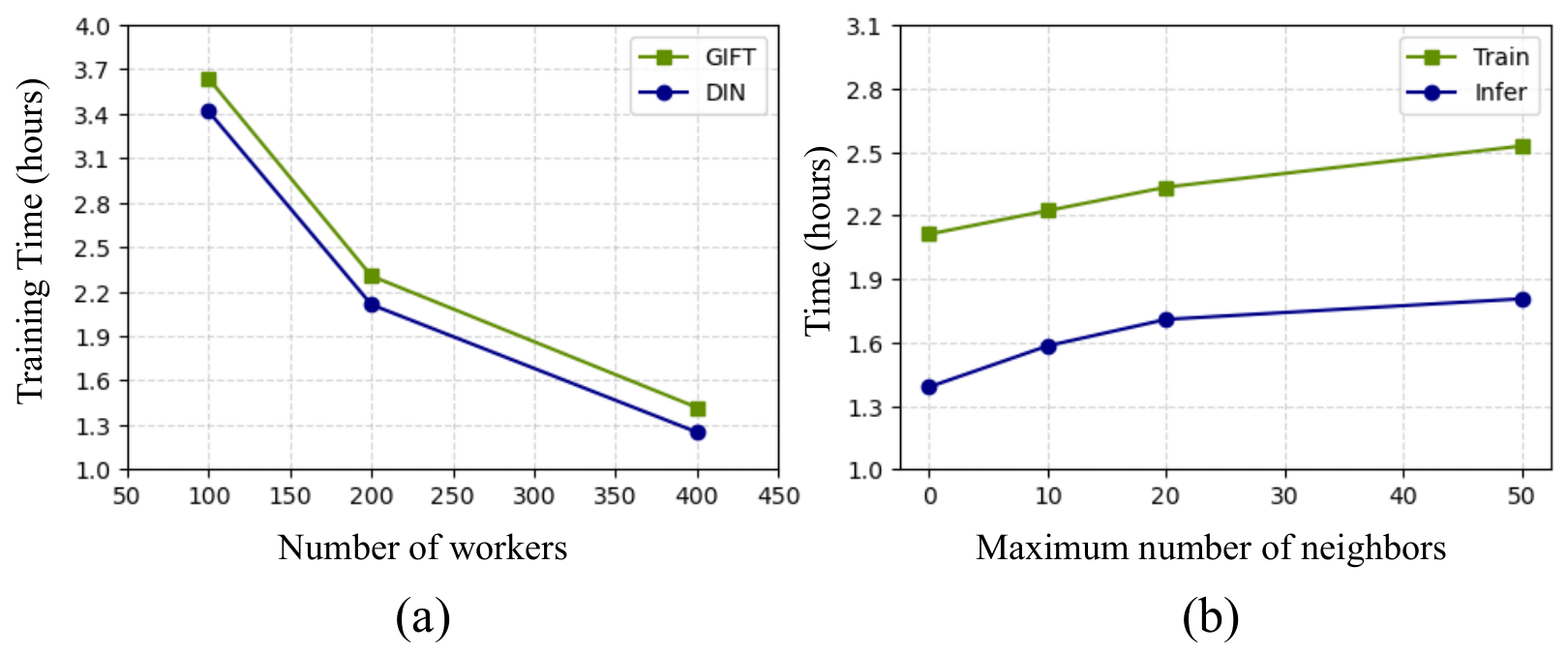}
  \caption{Study of Scalability. (a) presents the training overhead \textit{w.r.t.} the number of workers; (b) presents the scalability test \textit{w.r.t.} the maximum number of neighbors}
  \label{fig:scalability}
  \end{minipage}
\end{figure}

First, we investigate the scalability of GIFT that has been deployed on multiple workers for optimization. Figure~\ref{fig:scalability}(a) shows the training time of 1 million steps \textit{w.r.t.} the number of workers on Taobao dataset. The figure shows that GIFT is quite scalable on the distributed platform, as the training time decreases significantly when adding up the number of workers. Finally, when the number of workers reaches 400, the training speed of GIFT is very close to that of DIN, which indicates GIFT is scalable enough to be adopted in practice.
Second, we investigate the scalability of GIFT \textit{w.r.t.} the maximum number of neighbors for each cold-start video. As presented in Figure~\ref{fig:scalability}(b), the training time cost (green line) for 1 million steps has a linear growth at the beginning of the maximum number of neighbors scales up and then reaches a limit when it is larger than 50, this is because most cold-start videos cannot be linked to more than 50 videos as the frequency of node degree follows the power-law distribution. Similar is the trend of time overhead of inference on the Taobao dataset (blue line). The empirical evidence guarantees that GIFT does not require a relatively large number of neighbors to achieve good results, thus restricting the time overhead as well.



\section{Related Work}


\noindent \textbf{Content-based Methods:} content-based methods refer to the methods that exploit content information, such as item attributes, to address the cold-start dilemma. Content-based Filtering \cite{van2000using,thorat2015survey} is proposed with the assumption that if a user likes a product, it is very likely that she/he will prefer other attribute-similar items. By building this attribute relation, the content-based filtering is able to make cold-start item recommendation without requiring any behavior logs for new items. Along this line, different kinds of side information and algorithm are explored under different scenarios\cite{zhu2019addressing,gantner2010learning}.
    Another line of works is the hybrid models\cite{thorat2015survey,kim2006new} that take both cares of behavior logs and the content information. Take e-commerce recommendation\cite{EGES} for example, innate attributes of item like seller, category, brand, style etc are utilized as the initial feature for the item representation learning, makes their embeddings more robust and effective, despite the missing of user's behaviors for the cold-start items.

\noindent \textbf{Transfer learning Methods:} Transfer learning is widely used in the cold-start recommendation systems since it targets on applying knowledge extracted from the warmed-up items (source domain) to the cold-start items (target domain). Based on the transfer object, it can be roughly divided as instance- and model-based transfer learning. Instance-based transfer learning\cite{pan2008transfer,zhuang2011exploiting,liu2020heterogeneous} aims to extract useful information from the source domain data and convert them to the target domain with weighted tricks. Methods like feature mapping\cite{pan2008transfer}, data sharing\cite{zhuang2011exploiting} or building explicit relationships like graph structure\cite{liu2020heterogeneous} between two domains can be used to enhance the information of the cold-start items. It is noted that the graph-based transfer method elaborated in this paper falls into this category; Model-based transfer learning\cite{zhou2020s3,MMD,ganin2015unsupervised} aims to transfer model parameters trained on the abundant samples of source domain to the target domain meanwhile to prevent the negative transfer caused by the distribution gap. Training techniques like pre-training and fine-tuning\cite{zhou2020s3}, confusion regularization\cite{MMD}, adversarial training\cite{ganin2015unsupervised} are applied to make the model easily adapt to the target domain.


\section{Conclusion}
In this paper, we focus on the CTR prediction task for cold-start videos in Taobao.
We present an efficient graph-guided feature transfer system, GIFT, which establishes physical and semantic likages between warmed-up videos and cold-start ones, and transfer useful information to the target cold-start video guided by the graph.
When evaluated on a huge real-world dataset, GIFT achieves significantly better results than other competitors. Online A/B test further verifies its effectiveness for the recommendation task of the homepage of Taobao App. 

\section{Acknowledgement}
This work was partly supported by the Zhejiang Provincial Natural Science Foundation for Distinguished Young Scholars under No. LR19F020003, Zhejiang Lab under No.
2019KE0AB01, and the Alibaba-ZJU Joint Research Institute of Frontier Technologies.

\bibliographystyle{abbrv}
\bibliography{GIFT}

\end{document}